\documentclass[aps,prd,11pt,superscriptaddress,notitlepage,longbibliography,nofootinbib,tightenlines,groupedaddress]{revtex4-1}
\usepackage{amsmath,amssymb,amsfonts}
\usepackage{graphicx}
\usepackage{color}
\usepackage{tikz}
\usepackage{booktabs}
\usepackage{dsfont}
\usepackage[pdftex]{hyperref}
\hypersetup{colorlinks=true, linkcolor=darkred, citecolor=blue, linktoc=page}
\definecolor{darkred}{rgb}{0.8,0.1,0.1}

\usepackage{subfigure}

\newcommand{\N}[1]{\ensuremath{\mathcal N\,{=}\,#1}}

\def\cA{{\cal A}}
\def\cB{{\cal B}}
\def\cC{{\cal C}}
\def\cG{{\cal G}}

\DeclareMathOperator{\vol}{vol}
\DeclareMathOperator{\Vol}{Vol}

\DeclareMathOperator{\arcsinh}{arcsinh}
\newcommand{\bea}{\begin{eqnarray}}
\newcommand{\eea}{\end{eqnarray}}

\makeatletter\def\l@subsubsection#1#2{}%
\makeatother

\newcommand{\nocontentsline}[3]{}
\newcommand{\tocless}[2]{\bgroup\let\addcontentsline=\nocontentsline#1{#2}\egroup}

\begin{document}

\title{Entanglement entropy vs.\ free energy in IIB supergravity duals for 5d SCFTs}

\author{Michael Gutperle}
\email{gutperle@physics.ucla.edu}
\author{Chrysostomos Marasinou}
\email{cmarasinou@physics.ucla.edu}
\author{Andrea Trivella}
\email{andrea.trivella@physics.ucla.edu}
\author{Christoph F.~Uhlemann} 
\email{uhlemann@physics.ucla.edu}

\affiliation{Mani L.\ Bhaumik Institute for Theoretical Physics\\
Department of Physics and Astronomy\\
University of California, Los Angeles, CA 90095, USA}

\begin{abstract}
\baselineskip 16pt

We study entanglement entropy and the free energy in recently constructed holographic duals for 5d SCFTs in type IIB supergravity.
The solutions exhibit mild singularities, which could potentially complicate holographic applications. 
We use the relation of the entanglement entropy  
for a spherical entangling surface
to the free energy of the field theory  on the five sphere as a well-motivated benchmark to assess how problematic the  singularities are.
The holographic  supergravity computations give well-defined results for both quantities and they satisfy the expected relations.
This supports the interpretation of the solutions as holographic duals for 5d SCFTs and gives first quantitative indications for the nature of the dual SCFTs.
\end{abstract}

\maketitle
\newpage

\tableofcontents

\newpage

\baselineskip 16pt
\section{Introduction}

Five-dimensional superconformal field theories (SCFTs) are interesting for a variety of reasons. 
Their existence is not obvious, since Yang-Mills theories in 5d have a dimensionful coupling constant and are 
non-renormalizable by power counting. Therefore they can not be treated consistently in perturbation theory.
Nevertheless, the classification of   \cite{Nahm:1977tg} states that there is a  unique superconformal  algebra    with 16 supercharges in five dimensions, given by the superalgebra $F(4)$ \cite{Kac:1977em}.
Field theory analysis of the dynamics on the Coulomb branch indeed indicates that for large classes of combinations of gauge group and matter content, 5d super Yang-Mills theories admit a well-defined UV limit where the coupling constant diverges  \cite{Seiberg:1996bd,Intriligator:1997pq}.

There is no known standard Lagrangian description for the 5d SCFTs obtained as UV fixed points of the gauge theories.  However, the theories  can be engineered using brane constructions in type IIA and IIB string theory \cite{Aharony:1997ju,Aharony:1997bh,DeWolfe:1999hj}, which further supports their existence and has led to many insights.
In the absence of a conventional Lagrangian description, AdS/CFT dualities are a perfect tool for comprehensive quantitative analysis.
Supergravity duals in type IIA supergravity have indeed been known for some time, but they are singular \cite{Brandhuber:1999np,Bergman:2012kr}.
Although the singular nature limits the kind of questions that can be addressed, remarkable checks have been carried out with these solutions. In \cite{Jafferis:2012iv} the free energy on S$^5$ was compared to a localization calculation in the putative dual field theory.
Due to the singularities in the solutions, the holographic computation of the free energy had to proceed through the entanglement entropy of a spherical region, which is less sensitive to the singularities in the geometry.\footnote{
Another strategy is to work in 6d gauged supergravity, where the singularities resulting from the brane construction in type IIA string theory are not visible \cite{Assel:2012nf,Alday:2014rxa,Alday:2014bta}.
}
But once obtained, it matched the localization calculation, lending strong support to the proposed holographic dualities.

More recently, large classes of holographic duals for 5d SCFTs have been constructed in type IIB supergravity \cite{DHoker:2016ujz,DHoker:2016ysh,DHoker:2017mds}, where the geometry takes the form of AdS$_6\times$S$^2$ warped over a two-dimensional Riemann surface $\Sigma$.\footnote{
For earlier work on $AdS_6$ type IIB solutions  see \cite{Lozano:2012au,Lozano:2013oma,Apruzzi:2014qva,Kim:2015hya,Kim:2016rhs}.}
The solutions are singular as well, and  avoid a recent no-go theorem \cite{Gutowski:2017edr}.
But in contrast to the type IIA solutions, the singularities are at isolated points which have a clear interpretation as remnants of  the external 5-branes appearing in the brane-web constructions.
Nevertheless, it is an interesting question whether or to what extent these singularities affect AdS/CFT computations.

In this paper we study the holographic computation of the  free energy and entanglement entropy for the supergravity solutions of \cite{DHoker:2016ujz,DHoker:2016ysh,DHoker:2017mds}.  In particular, we calculate the finite part of the entanglement entropy for a spherical region and the free energy  of the field theory on S$^5$.
In view of the previous computations in type IIA supergravity, this is a well-motivated benchmark for how substantial the singularities are,
and it provides valuable quantitative information about the dual SCFTs.
We will show that the singularities are indeed mild enough to not interfere with the computation of either the free energy or the entanglement entropy. In fact, it appears that the poles also do not contribute a finite part in either calculation, which would be well in line with the interpretation that modes on the external 5-branes in brane web constructions decouple.
Moreover, the relation of the finite part of the entanglement entropy for a spherical region  to the free energy on S$^5$ holds as expected on general grounds \cite{Casini:2011kv}.  For the non-trivial geometries we are considering the  equivalence depends on rather non-trivial identities and hence provides a strong consistency check.

The rest of the paper is organized as follows. In sec.~\ref{sec:review} we briefly review the structure of the type IIB supergravity solutions \cite{DHoker:2016ujz,DHoker:2016ysh,DHoker:2017mds} and introduce the quantities that will be relevant for our computations. In sec.~\ref{sec:on-shell-action} we derive a general expression for the on-shell action of the solutions and discuss as special cases the 3- and 4-pole solutions that were spelled out in detail in \cite{DHoker:2017mds}.
This  provides a holographic calculation of  the free energy on S$^5$ for the dual SCFTs.
In sec.~\ref{sec:EE} we similarly discuss the computation of codimension-2 minimal surfaces anchored on the boundary of AdS$_6$ at a constant time, which compute the entanglement entropy   for the dual SCFTs. After deriving a general expression we discuss the same special cases as previously for the free energy, and show that the finite part for a spherical entangling region agrees with the free energy on S$^5$.
We close with a discussion of what our results imply for the nature of the dual SCFTs and directions for future research in sec.~\ref{sec:discussion}.

\section{Review of type IIB supergravity solutions}\label{sec:review}
The type IIB supergravity solutions we consider in this paper have been derived and discussed in detail in \cite{DHoker:2016ujz,DHoker:2016ysh,DHoker:2017mds},
and we will only give a brief review introducing the quantities  that will be relevant for the computation of 
free energy and entanglement entropy.

The relevant bosonic fields of type IIB supergravity are the metric, the complex axion-dilaton scalar $B$ and
the complex 2-form $C_{(2)}$ \cite{Schwarz:1983qr,Howe:1983sra}. The real 4-form $C_{(4)}$ and the fermionic fields vanish.
The geometry of the solutions is AdS$_6\times$S$^2$ warped over a Riemann surface $\Sigma$,
which for the solutions considered here will be the upper half plane.
With a complex coordinate $w$ on $\Sigma$, the metric and the 2-form field are parametrized by scalar functions $f_2^2$, $f_6^2$, $\rho^2$ and $\cC$ on $\Sigma$,
\begin{align}\label{eqn:ansatz}
 ds^2 &= f_6^2 \, ds^2 _{\mathrm{AdS}_6} + f_2 ^2 \, ds^2 _{\mathrm{S}^2} + 4\rho^2 dw d\bar w~,
 &
 C_{(2)}&=\cC \vol_{S^2}~.
\end{align}
The solutions are expressed in terms of two holomorphic functions $\cA_\pm$ on $\Sigma$, which are given by
\begin{align}\label{eqn:cA}
 \cA_\pm (w) &=\cA_\pm^0+\sum_{\ell=1}^L Z_\pm^\ell \ln(w-p_\ell)~.
\end{align}
The $p_\ell$ are restricted to be on the real line and are poles with residues $Z_\pm^\ell$ in $\partial_w \cA_\pm$. 
The residues are related by complex conjugation  $Z_\pm^\ell= - \overline{Z_\mp^\ell}$.
The explicit form of the solutions is conveniently expressed in terms of the composite quantities
\begin{align}\label{eq:kappa-G}
 \kappa^2&=-|\partial_w \cA_+|^2+|\partial_w \cA_-|^2~,
 &
 \partial_w\cB&=\cA_+\partial_w \cA_- - \cA_-\partial_w\cA_+~,
 \\
 \cG&=|\cA_+|^2-|\cA_-|^2+\cB+\bar{\cB}~,
 &
  R+\frac{1}{R}&=2+6\,\frac{\kappa^2 \, \cG }{|\partial_w\cG|^2}~.
  \label{eqn:Gdef}
\end{align}
Regularity of the solutions requires that $\kappa^2$ and $\cG$ are both positive in the interior of $\Sigma$ and vanish on the boundary.
These regularity conditions are satisfied if the residues  are given by
\begin{align}\label{eqn:residues}
Z_+^\ell  &=
 \sigma\prod_{n=1}^{L-2}(p_\ell-s_n)\prod_{k \neq\ell}^L\frac{1}{p_\ell-p_k}~.
\end{align}
and the $s_n$ are restricted to be in the upper half plane.
Moreover, the $p_\ell$ and $s_n$ have to be chosen such that they satisfy
\begin{align}
\label{eqn:constr}
 \cA^0 Z_-^k + \bar \cA^0 Z_+^k 
+ \sum _{\ell \not= k }Z^{[\ell k]} \ln |p_\ell - p_k| &=0~,
\end{align}
where $Z^{[\ell k]}\equiv Z_+^\ell Z_-^k-Z_+^k Z_-^\ell$ and $2\cA^0\equiv\cA^0_+-\bar\cA_-^0$.
The explicit form of the functions parametrizing the metric is then given by
\begin{align}\label{eqn:metric}
f_6^2&=\sqrt{6\cG} \left ( \frac{1+R}{1-R} \right ) ^{1/2}~,
&
f_2^2&=\frac{1}{9}\sqrt{6\cG} \left ( \frac{1-R}{1+R} \right ) ^{3/2}~,
&
\rho^2&=\frac{\kappa^2}{\sqrt{6\cG}} \left (\frac{1+R}{1-R} \right ) ^{1/2}~,
\end{align}
where we used the expressions of \cite{DHoker:2017mds} with $c_6^2=1$, which was shown there to be required for regularity.
The function $\cC$ parametrizing the 2-form field is given by
\begin{align}\label{eqn:flux}
 \cC = \frac{4 i }{9}\left (  
\frac{\partial_{\bar w} \bar \cA_- \, \partial_w \cG}{\kappa ^2} 
- 2 R \, \frac{  \partial_w \cG \, \partial_{\bar w} \bar \cA_- +  \partial_{\bar w}  \cG \, \partial_w \cA_+}{(R+1)^2 \, \kappa^2 }  
 - \bar  \cA_- - 2 \cA_+ \right )
\end{align}
and the axion-dilaton scalar $B$ is given by 
\begin{align}
B &=\frac{\partial_w \cA_+ \,  \partial_{\bar w} \cG - R \, \partial_{\bar w} \bar \cA_-   \partial_w \cG}{
R \, \partial_{\bar w}  \bar \cA_+ \partial_w \cG - \partial_w \cA_- \partial_{\bar w}  \cG}~.
\end{align}

\section{On-shell action and free energy on \texorpdfstring{S$^5$}{S5}}\label{sec:on-shell-action}

We will now evaluate the on-shell action for the solutions reviewed in the previous section explicitly.
Formulating an action for type IIB supergravity is subtle due to the self-duality constraint on the 4-form potential,
but since $C_{(4)}=0$ in our solutions this is not an issue.
Moreover, the on-shell action can be expressed as a boundary term \cite{Okuda:2008px}.
We relegate the details of translating the result of \cite{Okuda:2008px} to our convention to appendix \ref{app:action-boundary-term},
and start from the result (\ref{finresappa})
\begin{align}
 S_\mathrm{IIB}^\mathrm{E} &= 
 \frac{1}{64\pi G_\mathrm{N}}\int_{\mathcal M} d \Bigg[
 \frac{1}{2}f^2(1+|B|^2)\,\bar C_2\wedge \star d C_2
 -f^2 \bar B C_2\wedge \star d C_2+\mathrm{c.c.} \Bigg]
 \nonumber\\
 &=
 \frac{1}{64\pi G_\mathrm{N}}\int_{\partial \mathcal M}f^2\left[\frac{1}{2}(1+|B|^2)\bar C_2-\bar B C_2\right]\wedge \star d C_2+\mathrm{c.c.}
\end{align}
where $f^{-2}=1-|B|^2$.
We now use that $C_2=\mathcal C \mathrm{vol}_{\mathrm{S}^2}$, where $\mathrm{vol}_{\mathrm{S}^2}$ is the volume form on the S$^2$ of unit radius. This yields
\begin{align}
 \star d C_2&= f_6^6 f_2^{-2}\mathrm{vol}_{\mathrm{AdS}_6}\wedge \star^{}_\Sigma d\mathcal C~,
\end{align}
where $\mathrm{vol}_{\mathrm{AdS}_6}$ is the volume form on AdS$_6$ of unit curvature radius and $\star^{}_\Sigma$ is the Hodge dual on $\Sigma$ with metric $g_\Sigma=4\rho^2 |dw|^2$.
We then find
\begin{align}
  S_\mathrm{IIB}^\mathrm{E} &= \frac{1}{64\pi G_\mathrm{N}}\int_{\partial M}f^2 f_6^6 f_2^{-2}\left[\frac{1}{2}(1+|B|^2)\bar{\mathcal C}-\bar B \mathcal C\right]\mathrm{vol}_{\mathrm{S}^2}\wedge \mathrm{vol}_{\mathrm{AdS}_6}\wedge \star^{}_\Sigma d\mathcal C + \mathrm{c.c.}
\end{align}
The AdS$_6$ volume can be regularized and renormalized in the usual way for an AdS$_6$ with unit radius of curvature
and we will just use $\mathrm{Vol}_{\mathrm{AdS}_6,\mathrm{ren}}$ to denote the renormalized volume.
As we discuss in appendix \ref{sec:ads6vol}, there are no finite contributions to the on-shell action from the boundary introduced when regularizing the AdS$_6$ volume.
The explicit expression for the renormalized volume of global AdS$_6$ with a renormalization scheme preserving the S$^5$ isometries of the sphere slices is also derived in appendix \ref{sec:ads6vol} and given by
\begin{align}
\mathrm{Vol}_{\mathrm{AdS}_6,\mathrm{ren}}=-\frac{8}{15}\mathrm{Vol}_{\mathrm{S}^5}~.
\end{align}
Note that we denote by e.g.\ $\Vol_{\mathrm{S}^5}$ the actual volume, i.e.\ $\Vol_{\mathrm{S}^5}=\int_{\mathrm{S}^5}\vol_{\mathrm{S}^5}$.
The only (remaining) boundary then is the boundary of $\Sigma$. 
We note that $\partial\Sigma$ is not an actual boundary of the ten-dimensional geometry, so in particular there are no extra boundary terms to be added, but for the evaluation of the on-shell action as a total derivative we have to take it into account.
We thus find
\begin{align}\label{eq:on-shell0}
 S_\mathrm{IIB}^\mathrm{E} &=\frac{1}{64\pi G_\mathrm{N}}\mathrm{Vol}_{\mathrm{AdS}_6,\mathrm{ren}} \mathrm{Vol}_{\mathrm{S}^2} \int_{\partial \Sigma}f^2 f_6^6 f_2^{-2}\left[\frac{1}{2}(1+|B|^2)\bar{\mathcal C}-\bar B \mathcal C\right] \star_\Sigma d\mathcal C + \mathrm{c.c.}
\end{align}
The task at hand is to evaluate the various ingredients in this expression more explicitly.
To evaluate the metric factors more explicitly we use the expressions in (\ref{eqn:metric}), which yields
\begin{align}\label{eq:f6f2}
 f_6^6 f_2^{-2}&=54  \,\cG \left(\frac{1+R}{1-R}\right)^3~.
\end{align}
The pullback of  $\star_\Sigma d\mathcal C$ to $\partial\Sigma$ does not involve $\rho^2$, and to evaluate it explicitly we note that $\partial\Sigma=\mathds{R}$. 
It will be convenient for the explicit expansions to introduce real coordinates, $w=x+iy$, which yields
\begin{align}\label{eq:starC}
 \star_\Sigma d\mathcal C&=-(\partial_y\mathcal C)  dx~.
\end{align}
Using eq.~(\ref{eq:f6f2}) and (\ref{eq:starC}), the regularized on-shell action (\ref{eq:on-shell0}) becomes
\begin{align}
 S_\mathrm{IIB}^\mathrm{E} &=-\frac{1}{64\pi G_\mathrm{N}}54  \mathrm{Vol}_{\mathrm{AdS}_6,\mathrm{ren}} \mathrm{Vol}_{\mathrm{S}^2} 
 \int_{\mathds{R}} dx\, f^2 \cG \left(\frac{1+R}{1-R}\right)^3(\partial_y \mathcal C)
 \left(\frac{1}{2}(1+|B|^2)\bar {\mathcal C} -\bar B \mathcal C\right)+\mathrm{c.c.}\,,
\end{align}
where the integrand is evaluated at $y=0$.
Close to the boundary we have $\kappa^2,\cG\rightarrow 0$ and
\begin{align}
 R&=1-\sqrt{\frac{6\kappa^2\cG}{|\partial_w\cG|^2}}+\ldots~.
\end{align}
As discussed in sec.~5.5 of \cite{DHoker:2016ujz}, $\cG/(1-R)$ remains finite at the boundary
and the same applies for $f^2$.
We can thus simplify the on-shell action to
\begin{subequations}\label{eq:int}
\begin{align}
 S_\mathrm{IIB}^\mathrm{E} &= \frac{1}{8\pi G_\mathrm{N}}\mathrm{Vol}_{\mathrm{AdS}_6,\mathrm{ren}} \mathrm{Vol}_{\mathrm{S}^2} I_0~,
 \label{eq:int-1}
 \\
 I_0&=
 54 
 \int_{\mathds{R}} dx\, \frac{\cG}{1-R}\times\frac{ \partial_y \mathcal C}{(1-R)^2}\times
 \left(\bar B f^2 \mathcal C-\frac{2f^2-1}{2}\bar {\mathcal C}\right)+\mathrm{c.c.}~,
\end{align}
\end{subequations}
where each factor in the integrand is finite separately on the real line.

\subsection{Explicit expansions}
To further evaluate the on-shell action in (\ref{eq:int}), we explicitly expand the composite quantities $\kappa^2$, $\cG$ 
as well as the actual supergravity fields around the real line, and it turns out that the subleading orders in the expansion play a crucial role.
For the explicit expansions it is convenient to introduce
\begin{align}\label{eq:fpmdpm}
 f_\pm &= \cA_\pm^0+\sum_{\ell=1}^L Z_\pm^\ell \ln | x-p_\ell|~,
 &
 D_\pm&=i\pi \sum_{\ell=1}^LZ_\pm^\ell \Theta(p_\ell-x)~,
\end{align}
such that the holomorphic functions $\cA_\pm$ and their differentials can be written as
\begin{align}
 \cA_\pm&=D_\pm + \sum_{n=0}^\infty \frac{1}{n!}(iy)^n f^{(n)}_\pm~,
 &
 \partial_w\cA_\pm&=\sum_{n=0}^\infty \frac{1}{n!}(iy)^n f^{(n+1)}_\pm~,
\end{align}
where $f^{(n)}_\pm= (\partial_x)^n f_\pm$.
The composite quantity $\kappa^2$ can be evaluated straightforwardly. 
For $\cG$ we use $\partial_y \cG=i(\partial_w\cG-\partial_{\bar w}\cG)$ along with the fact that $\cG=0$ on the boundary.
This allows us to simply integrate the explicit expression for $\partial_y\cG$, which can be obtained straightforwardly from (\ref{eq:kappa-G}), to obtain an explicit expression for the expansion of $\cG$ in $y$.
We then find
\begin{align}
 \kappa^2&=y\kappa_0^2+\frac{1}{6}y^3\kappa_3^2+\mathcal O(y^5)~,\\
 \cG&=y\cG_0+\frac{1}{6}y^3\cG_3+\mathcal O(y^5)~,
\end{align}
where
\begin{align}
 \kappa_0^2&=2i(f_-^\prime f_+^{\prime\prime}-f_-^{\prime\prime} f_+^\prime)~,
 &
 \kappa_3^2&=-(\kappa_0^2)^{\prime\prime}+8i\left(f_+^{\prime\prime\prime}f_-^{\prime\prime}-f_-^{\prime\prime\prime} f_+^{\prime\prime}\right)~,
 \\
 \cG_0&=
 4 i \left(f_+ f_-^\prime-f_- f_+^\prime\right)~,
 &
 \cG_3&=-(\cG_0)^{\prime\prime}-4\kappa_0^2~.
\end{align}
The expansion coefficients are real by construction and, by the regularity conditions, $\kappa_0^2>0$ and $\cG_0>0$.
Since $\cG$ is constant along each piece of the boundary without poles, 
we also have $|\partial_w\cG|^2=\frac{1}{4}|\partial_y\cG|^2=\frac{1}{4}\cG_0^2$ 
(noting that $\partial_w=\frac{1}{2}(\partial_x-i\partial_y)$).
Using these expansions to find $\cC$ yields
\begin{align}
 \cC&=-\frac{4}{3}i D_+ + \frac{4}{9}y^3\left[\frac{f_+^\prime\cG_3}{\cG_0}+f_+^{\prime\prime\prime}+\frac{6\kappa_0^2}{\cG_0^2}\left(3f_+^\prime\cG_0-f_+(\cG_0)^\prime\right)\right]~.
\end{align}
This shows that the factors in (\ref{eq:int}) are indeed all finite as $y\rightarrow 0$.
The last ingredient we need is the limit  of $B$ at  the real axis, for which we find
\begin{align}
 B&=\frac{2f_+\kappa_0 - if_+^\prime\sqrt{6\cG_0}}{if_-^\prime\sqrt{6\cG_0}-2f_-\kappa_0}~,
\end{align}
and we note that this is not a pure phase.
Finally, for $f^2$ this yields
\begin{align}
 f^2&=\frac{1}{2}-\frac{4f_+f_-\kappa_0^2+6f_+^\prime f_-^\prime\cG_0}{\sqrt{6\cG_0^3\kappa_0^2}}~.
\end{align}

With the explicit expansions in hand, we now return to evaluating the integral $I_0$ in (\ref{eq:int}).
For the factors in the integrand we find
\begin{align}
 \frac{\cG}{1-R}\times \frac{\partial_y\cC}{(1-R)^2}&=
 \sqrt{\frac{\cG_0^3}{24\kappa_0^2}}\times
 \frac{1}{18\kappa_0^2}\left[f_+^\prime \cG_3+\cG_0 f_+^{\prime\prime\prime}+6\kappa_0^2\left(3f_+^\prime-f_+(\ln\cG_0)^\prime\right)\right]
 ~,\\
 \bar B f^2 \mathcal C-\frac{2f^2-1}{2}\bar {\mathcal C}&=
 -\frac{4}{3}\frac{i}{\sqrt{6\cG_0^3\kappa_0^2}}\left[
 6\cG_0 f_-^\prime\left(D_+f_-^\prime-D_- f_+^\prime\right)+4\kappa_0^2f_-\left(D_+ f_- - D_- f_+\right)
 \right]~,
\end{align}
where we used $\bar B f^2=(6\cG_0 (f_-^\prime)^2+4\kappa_0^2f_-^2)/\sqrt{6\cG_0^3\kappa_0^2}$ for the last expression.
The full integral then becomes
\begin{align}
 I_0&=-\frac{i}{3} \int_{\mathds{R}}dx
 \frac{1}{\kappa_0^4}
 \left[f_+^\prime \cG_3+\cG_0 f_+^{\prime\prime\prime}+6\kappa_0^2\left(3f_+^\prime-f_+(\ln\cG_0)^\prime\right)\right]
 \times
 \nonumber\\&\hskip 2.0in
 \left[
 6\cG_0 f_-^\prime\left(D_+f_-^\prime-D_- f_+^\prime\right)+4\kappa_0^2f_-\left(D_+ f_- - D_- f_+\right)
 \right]+\mathrm{c.c.}
\end{align}
Adding the complex conjugate explicitly yields
\begin{align}\label{eq:I0}
 I_0&=\int_{\mathds{R}}dx\,
 \cG_0
 \left[
 \frac{16}{3}(D_+f_--D_-f_+)
 -\frac{\cG_0(\kappa_0^2)^\prime-3(\cG_0)^\prime\kappa_0^2}{\kappa_0^4}(D_+ f^\prime_--D_- f_+^\prime)
 \right]~.
\end{align}
Via (\ref{eq:int}), this translates to an explicit expression for the on-shell action.

\subsection{Integrability of the poles}
We will now show that the integrand in (\ref{eq:I0}) is well-behaved at the poles, $x=p_\ell$, such that the integral can be evaluated straightforwardly.
To this end, we first evaluate $\cG_0$ and $\kappa_0^2$ more explicitly.
For $\cG_0$ we find, by straightforward evaluation,
\begin{align}
 \cG_0&=4i\sum_{k=1}^L\frac{\cA_+^0 Z_-^k-\cA_-^0 Z_+^k}{x-p_k}+4i\sum_{\ell\neq k}Z^{[\ell k]}\frac{\ln |x-p_\ell|}{x-p_k}~.
\end{align}
The integration constants $\cA_\pm^0$ are constrained by the regularity conditions  (\ref{eqn:constr}),
which, with $\cA_+^0=-\bar\cA_-^0$, read
\begin{align}
\label{eqn:constr2}
 \cA_+^0 Z_-^k - \cA_-^0 Z_+^k 
+ \sum _{\ell \not= k }Z^{[\ell k]} \ln |p_\ell - p_k| &=0~.
\end{align}
We therefore find that for generic solutions satisfying the regularity conditions 
\begin{align}\label{eqn:G0expl}
 \cG_0&=4i\sum_{k=1}^L\sum_{\ell\neq k}\frac{Z^{[\ell k]}}{x-p_k}\ln\left|\frac{x-p_\ell}{p_\ell-p_k}\right|~.
\end{align}
The evaluation of $\kappa_0^2$ is straightforward and yields
\begin{align}
 \kappa_0^2&=2i\sum_{k=1}^L\sum_{\ell\neq k}\frac{Z^{[\ell k]}}{(x-p_\ell)(x-p_k)^2}~.
\end{align}
Moreover, due to the antisymmetry of $Z^{[\ell k]}$ the derivatives of $\cG_0$ and $\kappa_0^2$  take a simple form and are given by
\begin{align}
 (\cG_0)^\prime&=-4i\sum_{k=1}^L\sum_{\ell\neq k}\frac{Z^{[\ell k]}}{(x-p_k)^2}\ln\left|\frac{x-p_\ell}{p_\ell-p_k}\right|~,
 &
 (\kappa_0^2)^\prime&=-4i\sum_{k=1}^L\sum_{\ell\neq k}\frac{Z^{[\ell k]}}{(x-p_\ell)(x-p_k)^3}~.
\end{align}

With these expressions in hand, we can now analyze the behavior of the integrand in (\ref{eq:I0}).
We set $x=p_m+\epsilon$, where $\epsilon$ is real and $|\epsilon|$ small compared to $1$ and to all $|p_k-p_\ell|$,
and find
\begin{align}
 \cG_0&=4i\eta_m\ln |\epsilon|+\mathcal O(1)~,
 &
 \kappa_0^2&=-2i\frac{\eta_m}{\epsilon^2}+\mathcal O(\epsilon^{-1})~,
 &
 \eta_m&=\sum_{k\neq m}\frac{Z^{[m k]}}{p_m-p_k}~,
 \\
 (\cG_0)^\prime&=\mathcal O(\epsilon^{-1})~,
 &
 (\kappa_0^2)^\prime&=4i\frac{\eta_m}{\epsilon^3}+\mathcal O(\epsilon^{-1})~.
\end{align}
Note that the behavior of $\cG_0$ and $(\cG_0)^\prime$ would be different if the parameters were not constrained by the regularity conditions in~(\ref{eqn:constr}).
The near-pole expansions consequently would be qualitatively different.
For regular solutions, however, it is now straightforward to verify,
with the explicit expansions of the composite quantities around the pole, that the integrand in~(\ref{eq:I0}) is $\mathcal O\big((\ln |\epsilon|)^2\big)$ and thus integrable across the pole.

\subsection{The on-shell action}\label{subsec:on-shell-action integral}
The integral $I_0$ in (\ref{eq:I0}) can be further simplified as follows.
We isolate the second term in the square brackets and rewrite the sum in the numerator as a total derivative,
\begin{align}\label{eq:I0a}
 I_0&=I_1+\frac{16}{3}\int_{\mathds{R}}dx\, \cG_0(D_+f_--D_-f_+)~,
 &
 I_1&=\int_{\mathds{R}}dx\,
 \bigg(\frac{\cG_0^3}{\kappa_0^2}\bigg)^\prime \; \frac{D_+ f^\prime_--D_- f_+^\prime}{\cG_0}~.
\end{align}
Using integration by parts we can further evaluate $I_1$.
This yields
\begin{align}
 I_1&=\frac{\cG_0^2}{\kappa_0^2}(D_+f_-'-D_-f_+')\bigg\vert_{x=-\infty}^{x=\infty}
 -\int_{\mathds{R}}dx\,\frac{\cG_0^3}{\kappa_0^2}\bigg(\frac{D_+ f^\prime_--D_- f_+^\prime}{\cG_0}\bigg)^\prime
 ~.
\end{align}
The first term vanishes, since $D_\pm=0$ if either $x>p_\ell$ or $x<p_\ell$ for all $\ell$,
thanks to $\sum_\ell Z_\pm^\ell=0$.
The $D_\pm$ given in (\ref{eq:fpmdpm}) depend on $x$ only through $\Theta$-functions, and we have to take into account their non-trivial  distributional derivatives.
The second term then evaluates to 
\begin{align}
 I_1&=
 -\int_{\mathds{R}}dx\,\Bigg[\frac{\cG_0^2}{\kappa_0^2}(D_+ f_-^{\prime\prime}-D_- f_+^{\prime\prime})+
 \frac{\cG_0^2}{\kappa_0^2}(D_+^\prime f_-^{\prime}-D_-^\prime f_+^{\prime})
 -\frac{\cG_0^\prime\cG_0}{\kappa_0^2}(D_+ f_-^{\prime}-D_- f_+^{\prime})
 \Bigg]~.
\end{align}
Since $D_\pm^\prime= - i\pi \sum_{\ell=1}^L Z_\pm^\ell \delta(p_\ell-x)$ and, by the analysis of the previous subsection,
$\cG_0^2f_\pm^\prime/\kappa_0^2=\mathcal O(\epsilon(\ln|\epsilon|)^2)$ close to the poles, the second term vanishes.
The first and last term can be combined thanks to the following identity, which follows from the expressions for $\kappa_0^2$ and $\cG_0$ in terms of $f_\pm$,
\begin{align}
 2f_\pm\kappa_0^2&=\cG_0f_\pm^{\prime\prime}-\cG_0^\prime f_\pm^\prime~.
\end{align}
The result is
\begin{align}
 I_1&=-2\int_{\mathds{R}}dx\,\cG_0(D_+f_--D_-f_+)~.
\end{align}
This reproduces exactly the structure of the remaining term in $I_0$ in (\ref{eq:I0a}) and we simply find
\begin{align}\label{eq:I0b}
 I_0&=\frac{10}{3}\int_{\mathds{R}}dx\, \cG_0(D_+f_--D_-f_+)~.
\end{align}
Evaluating $D_+f_--D_-f_+$ more explicitly, using the regularity condition (\ref{eqn:constr2}), yields
\begin{align}
 D_+f_--D_-f_+&=i\pi\sum_{k=1}^L\sum_{\ell\neq k}\Theta(p_\ell-x)Z^{[\ell k]}\ln\left|\frac{x-p_k}{p_\ell-p_k}\right|~.
\end{align}
Together with (\ref{eqn:G0expl}) this shows that $I_0$ explicitly depends on the residues only through the combinations $Z^{[\ell k]}$.
With (\ref{eq:int}), we finally find the on-shell action as
\begin{align}
\label{eq:on-shell-action}
 S_\mathrm{IIB}^\mathrm{E}&=-\frac{5 }{3 G_\mathrm{N}} \mathrm{Vol}_{\mathrm{AdS}_6,\mathrm{ren}} \mathrm{Vol}_{\mathrm{S}^2}
 \sum_{\substack{\ell,k,m,n=1 \\ \ell\neq k, m\neq n}}^L Z^{[\ell k]}Z^{[m n]}
 \int_{-\infty}^{p_\ell}dx\,\ln\left|\frac{x-p_k}{p_\ell-p_k}\right| \,\ln\left|\frac{x-p_m}{p_m-p_n}\right| \frac{1}{x-p_n}~.
\end{align}
We note that the lower bound in the integral can be moved from $-\infty$ to $\min_\ell (p_\ell)$ due to $\sum_\ell Z_+^\ell=0$.
The integral can be solved explicitly and involves polylogarithms.
While the result for generic configurations does not seem particularly illuminating, this allows us to get analytic results for particular solutions, as we will discuss in sec.~\ref{subsec:3-4-5 poles action}.
Note also that the $Z^{[\ell k]}$ are imaginary, so the expression (\ref{eq:on-shell-action}) is manifestly real.

\subsection{Scaling of the free energy}\label{sec:scaling}
As shown in \cite{DHoker:2017mds}, the residues $Z_\pm^\ell$  of the differentials $\partial_w\cA_\pm$ at the poles $p_\ell$ correspond to the charges of external 5-branes in brane-web constructions for 5d SCFTs. 
The details of the SCFT depend on the precise charge assignments, and the same applies for the free energy and, correspondingly, the gravitational on-shell action.
Before coming to those details, we can address a more general question: how does the free energy scale under overall rescalings of the 5-brane charges?

To address this question we can assume to start with a generic solution to the regularity conditions in (\ref{eqn:constr}). Namely,
\begin{align}
 \cA^0 Z_-^k + \bar \cA^0 Z_+^k 
+ \sum _{\ell \not= k }Z^{[\ell k]} \ln |p_\ell - p_k| &=0~.
\end{align}
We note that the equation is invariant under the following scaling
\begin{align}
 Z_+^\ell &\rightarrow \gamma Z_+^\ell~,
 &
 Z_-^\ell &\rightarrow \bar\gamma Z_-^\ell~,
 &
 \cA^0&\rightarrow \gamma \cA^0~,
 &
 p_\ell&\rightarrow p_\ell~,
\end{align}
where we have allowed for $\gamma\in\mathds{C}$.
For the residues this simply amounts to a change of the overall complex normalization parametrized by $\sigma$ in (\ref{eqn:cA}).
So starting with a solution $(Z_\pm^\ell,\cA^0,p_\ell)$ to the regularity conditions, a rescaling of this form produces another solution,
and this precisely allows us to isolate the overall scale of the charges $Z_+^\ell$.
From (\ref{eq:on-shell-action}) we immediately see that the on-shell action scales as
\begin{align}
 S_\mathrm{IIB}^\mathrm{E}\rightarrow |\gamma|^4 S_\mathrm{IIB}^\mathrm{E}~.
\end{align}

For a real overall scaling by $N$, we thus obtain a free energy scaling as $N^4$. 
This is different from the $N^2$ scaling one would expect for the 't Hooft limit of a four dimensional Yang-Mills theory, and as exhibited by \N{4} SYM and its AdS$_5\times$S$^5$ dual. But this is certainly not surprising, given the more exotic nature of the field theories described by 5-brane web constructions. It is also different from the $N^{5/2}$ scaling exhibited by the UV fixed points of 5d USp($N$) gauge theories and their gravity duals \cite{Jafferis:2012iv}. As a curious aside, however, we note that the free energy for the orbifold quivers obtained from the USp($N$) theories, which scales as $N^{5/2}k^{3/2}$, shows the same scaling if one na\"ively sets $k=N$.
As discussed in \cite{DHoker:2017mds}, there actually are classes of brane intersections described by the solutions discussed here which would naturally correspond to long quiver gauge theories with gauge groups of large rank, and we will discuss these examples in more detail in the next section.

\subsection{Solutions with 3, 4 and 5 poles}\label{subsec:3-4-5 poles action}
We now evaluate the general expression for the free energy in (\ref{eq:on-shell-action}) for classes of solutions with 3 up to 5 poles.
It will be convenient to separate off the general overall factors as in (\ref{eq:int-1}),
and focus on the solution-specific part $I_0$.

\subsubsection{3-pole solutions}
We start with the 3-pole case. As discussed in sec.~4.1 of \cite{DHoker:2017mds}, the SL($2,\mathds{R}$) automorphisms of the upper half plane can be used to fix the position of all poles, which we once again choose as 
\begin{align}\label{eq:3-poles-fixed}
 p_1&=1~, & p_2&=0~, & p_3&=-1~.
\end{align}
The regularity conditions are solved by $\cA^0=\omega_0\lambda_0 s \ln 2$.
The free parameters  of the solutions are given by the residues, corresponding to the charges of the external 5-branes, subject to charge conservation.
The integral $I_0$ in (\ref{eq:I0b}) for a generic choice of residues evaluates to
\begin{align}\label{eq:3-pole-I0}
 I_0&=-80\pi \zeta(3) (Z^{[12]})^2~.
\end{align}
The on-shell action therefore is a simple function that is quartic in the residues, and manifestly invariant under the SU$(1,1)$ duality symmetry of type IIB supergravity since the $Z^{[\ell k]}$ are.\footnote{%
The transformations spelled out in sec.~5.1 of \cite{DHoker:2016ujz} can be realized by transforming the residues as
$Z_+^\ell\rightarrow u Z_+^\ell-v Z_-^\ell$ and $Z_-^\ell\rightarrow \bar u Z_-^\ell-\bar v Z_+^\ell$.
}
Note also that $Z^{[\ell k]}$ is imaginary, and $I_0$ positive.
For the particular case of the ``$N$-junction'' \cite{Benini:2009gi}, discussed in sec.~4.3 of \cite{DHoker:2017mds} and realized by the charge assignment
$Z_+^1=N$, $Z_+^2=iN$, we have $Z^{[12]}=2iN^2$ and thus find the free energy quartic in $N$.

\subsubsection{4-pole solutions}

For solutions with four poles we can once again fix the position of three poles by SL($2,\mathds{R}$), but the position of one pole remains a genuine parameter. It is fixed by the regularity conditions in (\ref{eqn:constr}) and thus becomes an in general non-trivial function of the residues. We therefore expect in general more interesting dependence on the charges compared to the 3-pole case.
However, for the special class of 4-pole solutions discussed in sec.~4.2 of \cite{DHoker:2017mds}, where
\begin{align}\label{eq:4pole-special}
 Z_+^3&=-Z_+^1~,
 &
 Z_+^4&=-Z_+^2~,
\end{align}
the position of the fourth pole is independent of the residues.
In that case the regularity conditions are solved by
\begin{align}
 p_1&=1~, & p_2&=\frac{2}{3}~,& p_3&=\frac{1}{2}~,& p_4&=0~,
\end{align}
along with $\cA^0=Z_+^2\ln 3-Z_+^1\ln2$.
The position of all poles is therefore fixed regardless of the choice of charges, and we may again expect
the on-shell action to be a simple quartic function of the residues.
Indeed, the result for the integral is
\begin{align}\label{eq:4-pole}
 I_0&=-280\pi\zeta(3) (Z^{[12]})^2~,
\end{align}
and of the same general form as the 3-pole result. We also note the factor $\zeta(3)$ appearing again.
For the solutions discussed in sec.~4.2 of \cite{DHoker:2017mds}, with
$-Z_+^1=Z_+^3=(1+i)N$ and $Z_+^2=-Z_+^4=(1-i)M$, we have $Z^{[12]}=4iM N$.
In particular, for $M=N$ the free energy again scales like $N^4$, a feature which we will come back to in the discussion.

We will now discuss a different configuration with 4 poles, for which the position of the fourth pole actually depends on the choice of charges. To this end, it is convenient to move the position of one pole off to infinity, which we will discuss here for a generic $L$-pole solution.
To move the $L$-th pole $p_L$ to infinity, we perform the following replacements and limit
\begin{align}
 p_L&\rightarrow -\infty~,
 &
 \cA_\pm^0&\rightarrow \tilde \cA_\pm^0=\cA_\pm^0-Z_\pm^L\ln|p_L|~.
\end{align}
Note that the conjugation relation between the original integration 
constants, $\bar\cA_\pm^0=-\cA_\pm^0$, holds in the same form for 
$\tilde\cA_\pm^0$.
In terms of the redefined integration constants, the expressions for the holomorphic functions then become
\begin{align}
 \cA_\pm&=\tilde \cA_\pm^0+\sum_{\ell=1}^{L-1} Z_\pm^\ell \ln (w-p_\ell)~.
\end{align}
Note that this expression explicitly involves only $L-1$ poles and $L-1$ residues.
These residues, however, are not constrained to sum to zero and the number of independent parameters is therefore unchanged.
The conditions for $\cG=0$ on the boundary become
\begin{align}
 \tilde \cA_+^0 Z_-^k-\tilde \cA_-^0 Z_+^k+\sum_{\substack{\ell =1 \\ \ell\neq k}}^{L-1} Z^{[\ell k]}\ln |p_\ell-p_k|&=0~,
 &
 k&=1,..,L-1~.
\end{align}
These are only $L-1$ conditions, as compared to $L$ conditions previously. However, the sum does not manifestly vanish and the number of independent conditions therefore is also not modified.
The class of 4-pole solutions with (\ref{eq:4pole-special}) can now be realized as
\begin{align}
 p_1&=1~, & p_2&=0~, & p_3&=-1~,
 &
 \tilde\cA_\pm^0&=0~,
\end{align}
and computing the on-shell action reproduces (\ref{eq:4-pole}).

\begin{figure}[htb]
\center
\subfigure[][]{ \label{fig:theta-p3}
  \includegraphics[width=0.38\linewidth]{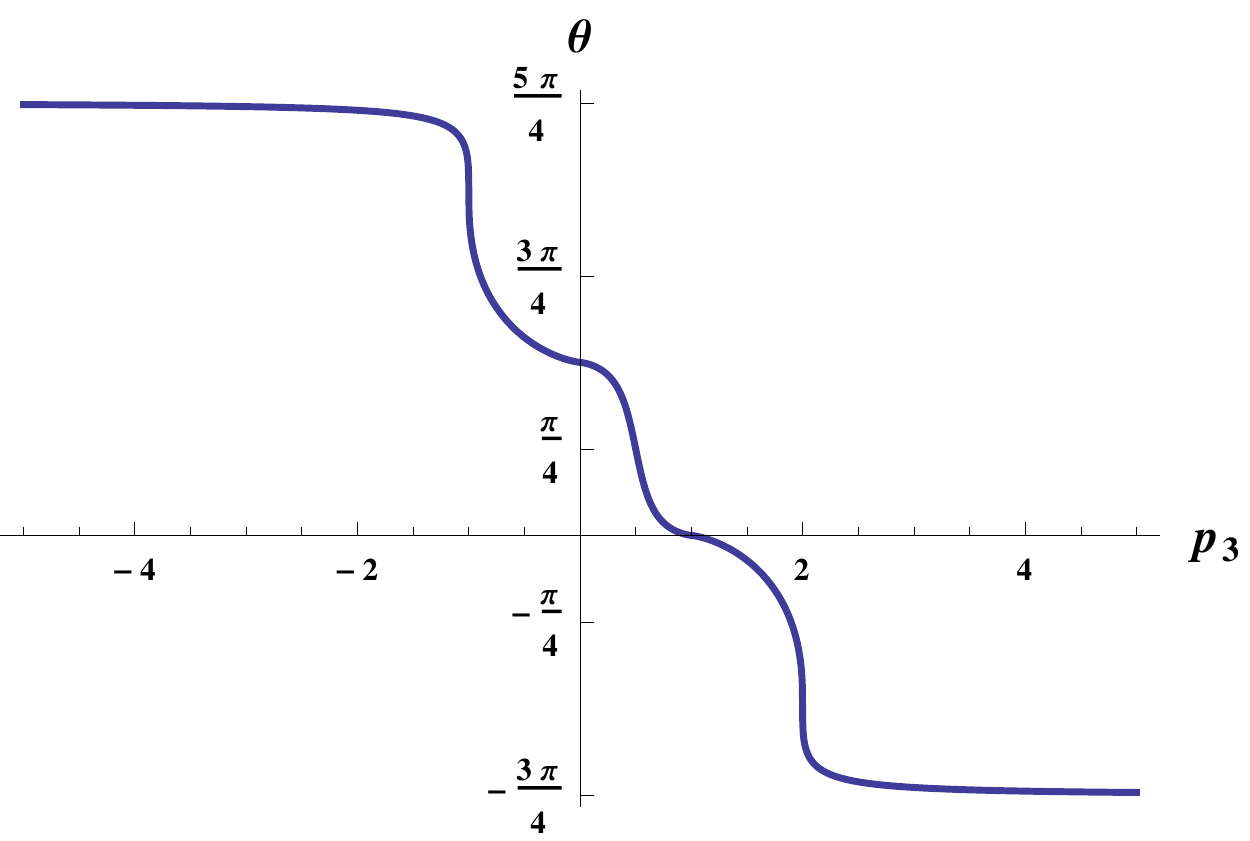}
}\hskip 0.5in
\subfigure[][]{ \label{fig:I0-theta}
 \includegraphics[width=0.38\linewidth]{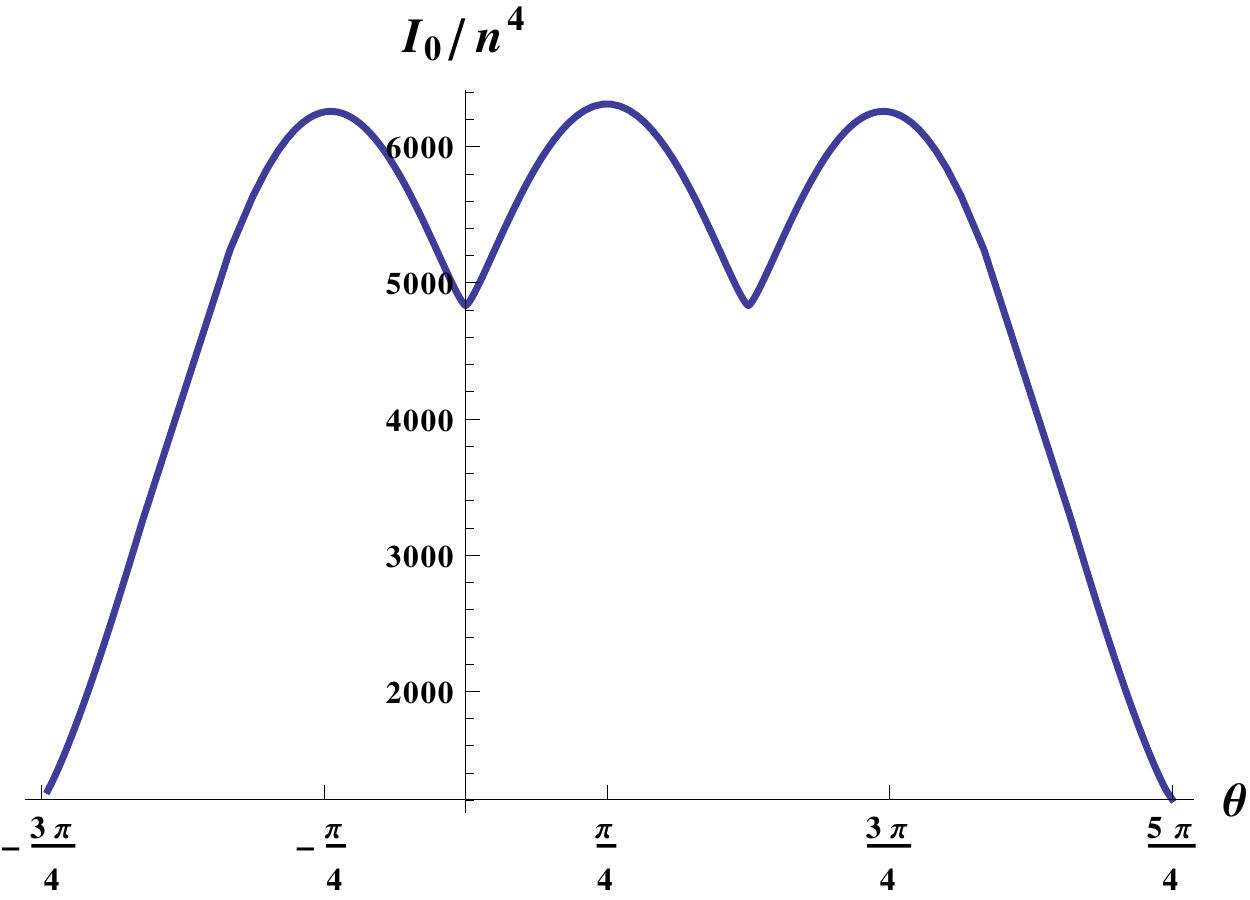}
}
\caption{The left hand side shows $\theta$ as function of $p_3$, for the 4-pole solution with residues given in eq.~(\ref{eq:4-pole-spin}). The right hand side shows $I_0$, which via (\ref{eq:int}) corresponds to the on-shell action.}
\end{figure}

The class of 4-pole solutions we wish to discuss next is parametrized by an overall scale $n$ of the residues and an angle $\theta$,
and obtained by fixing 
\begin{align}\label{eq:4-pole-spin}
 Z_+^1&=n~,  & Z_+^2&=in & Z_+^3&=n e^{i\theta}~, 
 &Z_+^4&=-(1+i+e^{i\theta})n~.
\end{align}
The position of three of the poles can once again be fixed arbitrarily, and we choose
\begin{align}
 p_1&=1~, & p_2&=0~, & p_4&\rightarrow -\infty~.
\end{align}
This leaves the position of the third pole, $p_3$, along with the (complex) constant $\cA^0$ to be determined from the conditions in (\ref{eqn:constr}). The resulting equation determining $p_3$ after solving for $\cA^0$ is
\begin{align}
 Z^{[1,3]}Z^{[2,4]}\ln(1-p_3)^2&= Z^{[1,4]}Z^{[2,3]}\ln p_3^2~.
\end{align}
Note that $n$ drops out of this equation and $p_3$ therefore depends on $\theta$ only.
We take the position of the pole as parameter and solve for $\theta$, which can be done in closed form and yields four branches of solutions.
The criterion for the choice of branch is that $\theta$ should be real and the zeros $s_n$ in the upper half plane.
The explicit expressions are bulky and not very illuminating, and we show a plot of $\theta$ as function of $p_3$ in fig.~\ref{fig:theta-p3} instead.
Since $p_3$ is independent of $n$, the on-shell action depends on $n$ only through an overall factor $n^4$, as expected from the scaling analysis in sec.~\ref{sec:scaling}.
The dependence on $\theta$, however, is non-trivial and we show the result in fig.~\ref{fig:I0-theta}.
We note the presence of three minima, which all correspond to the 4-pole solution degenerating to a 3-pole solution:
for $\theta\rightarrow 0$ we have $Z_+^3\rightarrow Z_+^1$ and $p_3\rightarrow p_1$,
for $\theta\rightarrow \pi/2$ we have $Z_+^3\rightarrow Z_+^2$ and $p_3\rightarrow p_2$,
and for $\theta\rightarrow 5\pi/4$ we have $Z_+^3\rightarrow (1+\sqrt{2})Z_+^4$ and $p_3\rightarrow p_4$.
That means in all these cases two poles coalesce and their residues add.
The free energy coincides with that of the resulting 3-pole configuration.
The 3-pole configurations resulting from $\theta\rightarrow 0$ and $\theta\rightarrow \pi/2$ have two charges with the same moduli and the same relative phase up to a sign. 
Since the formula in (\ref{eq:3-pole-I0}) is insensitive to these differences, this explains the coincident free energies.
It is intriguing to observe that the value of the free energy assumes a local minimum for all the cases where the solution reduces to a 3-pole configuration. The sphere free energy in odd dimension can be used as a measure for the number of degrees of freedom, and one may speculate that splitting one pole into two, or equivalently one external 5-brane into two, will generically increase that number. While certainly true for this specific example, it is an interesting open question whether this behavior holds more generally.

\subsubsection{5-pole solutions}

\begin{figure}[htb]
\center
\subfigure[][]{\label{fig:5-pole-a}
\tikzpicture[scale=0.9]

\draw (-0.16,-0.4) -- (-0.16,-2);
\draw (-0.08,-0.4) -- (-0.08,-2);
\draw (-0.0,-0.4) -- (-0.0,-2);
\draw (0.08,-0.4) -- (0.08,-2);
\draw (0.16,-0.4) -- (0.16,-2);

\draw (2,0.16) -- (0.4,0.16);
\draw (2,0.08) -- (0.4,0.08);
\draw (2,0.0) -- (0.4,0.0);
\draw (2,-0.08) -- (0.4,-0.08);
\draw (2,-0.16) -- (0.4,-0.16);

\draw (-0.16,0.4) -- (-0.16,2);
\draw (-0.08,0.4) -- (-0.08,2);
\draw (-0.0,0.4) -- (-0.0,2);
\draw (0.08,0.4) -- (0.08,2);
\draw (0.16,0.4) -- (0.16,2);

\draw[fill=gray] (0,0) circle (0.38);

\draw (-0.28,-0.28) -- (-1.41,-1.41);
\draw (-0.28-0.055,-0.28+0.055) -- (-1.41-0.056,-1.41+0.055);
\draw (-0.28-0.110,-0.28+0.110) -- (-1.41-0.112,-1.41+0.110);
\draw (-0.28+0.055,-0.28-0.055) -- (-1.41+0.056,-1.41-0.056);
\draw (-0.28+0.110,-0.28-0.110) -- (-1.41+0.112,-1.41-0.110);

\draw (-0.28,0.28) -- (-1.41,1.41);
\draw (-0.28-0.055,0.28-0.055) -- (-1.41-0.055,1.41-0.055);
\draw (-0.28-0.11,0.28-0.11) -- (-1.41-0.11,1.41-0.11);
\draw (-0.28+0.055,0.28+0.055) -- (-1.41+0.055,1.41+0.055);
\draw (-0.28+0.11,0.28+0.11) -- (-1.41+0.11,1.41+0.11);

\node at (-1.65,-1.65) {$N$};
\node at (-1.65,1.65) {$N$};
\node at (0,-2.3) {$M$};
\node at (2.35,0) {$2N$};
\node at (0,2.3) {$M$};

\endtikzpicture
}\hskip 0.8in
\subfigure[][]{\label{fig:5-pole-b}
 \includegraphics[width=0.42\linewidth]{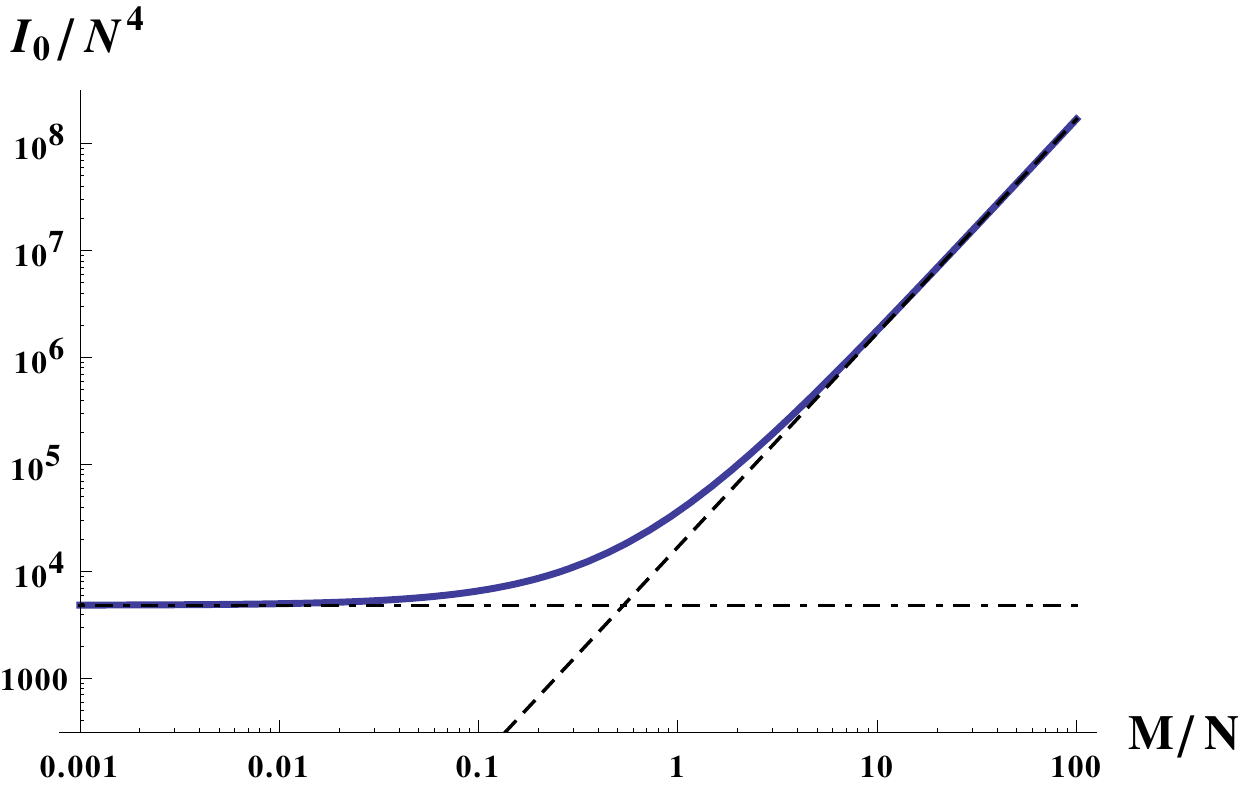}
}
\caption{
The left hand side shows a 5-brane intersection corresponding to the charges in (\ref{eqn:5pole-residues}).
On the right hand side is a $\log$-$\log$ plot of $I_0$ for the 5-pole solution with residues given in (\ref{eqn:5pole-residues}). 
Via (\ref{eq:int}) this corresponds to the on-shell action, as function of $M/N$.
The constant dot-dashed line shows $80\pi\zeta(3)\cdot 16N^4$, which, via (\ref{eq:3-pole-I0}), is the value of $I_0$ for the 3-pole solution resulting from (\ref{eqn:5pole-residues}) for $M=0$.
The dashed line shows $280\pi\zeta(3)\cdot 16M^2N^2$, which, via (\ref{eq:4-pole}), is $I_0$ for a 4-pole solution with 
$-Z_+^1=Z_+^3=2iN$ and $Z_+^2=-Z_+^4=M$.
\label{fig:5-pole}}
\end{figure}
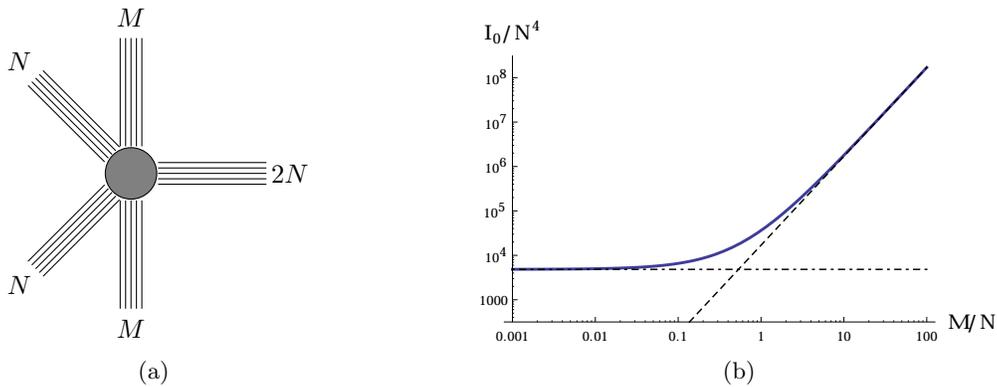

As a final example we will consider a class of solutions with five poles. 
In general we now have two positions of the poles depending on the choice of residues, 
but we will focus on a class of solutions which are parametrized by only two real numbers,
with residues given by 
\begin{align}\label{eqn:5pole-residues}
Z_+^1&=-Z_+^3=M~,
&
Z_+^2&=2iN~, & 
-Z_+^4&=iZ_+^5=(1+i)N~.
\end{align}
The corresponding 5-brane intersection is shown in fig.~\ref{fig:5-pole-a}.
As before three poles can be fixed by SL($2,\mathds{R}$) and we resort to the choice in (\ref{eq:3-poles-fixed}).
The regularity conditions in (\ref{eqn:constr}) are solved by
\begin{align}
 p_5&=-p_4~, & A^0&=iN\log|p_4^2-1|~,
\end{align}
where $p_4$ is determined by the equation
\begin{align}\label{eq:5pole-p4}
 (M-N)\log(p_4-1)^2-(M+N)\log(p_4+1)^2+N\log 16&=0~.
\end{align}
The choice of residues can be realized via (\ref{eqn:residues}),
by fixing $\sigma=-2iNp_4^2/(s_1s_2s_3)$ and the zeros $s_1$, $s_2$, $s_3$ as the three solutions to the cubic equation
\begin{align}
 isM(s^2-p_4^2)+p_4 N (s^2-1)(p_4-is)&=0~.
\end{align}
To solve (\ref{eq:5pole-p4}) it is once again convenient to fix $p_4$ and determine the resulting ratio $M/N$.
We choose $p_4\leq -\sqrt{5}$, which produces zeros in the upper half plane and positive $M/N$.
The on-shell action divided by $N^4$, as function of the ratio $M/N$, is shown in fig.~\ref{fig:5-pole-b}.
We clearly see that the dependence on $M/N$ is not simply quadratic, which we would have expected if the position of the poles had not depended on $M/N$.
Instead, $I_0/N^4$ interpolates between approaching a constant for small $M/N$ and quadratic dependence for large $M/N$.

\begin{figure}[htb]
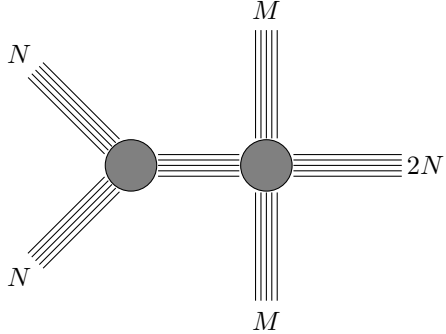

\center
\tikzpicture[scale=0.9]

\draw (2.0-0.16,-0.4) -- (2.0-0.16,-2);
\draw (2.0-0.08,-0.4) -- (2.0-0.08,-2);
\draw (2.0-0.0,-0.4) -- (2.0-0.0,-2);
\draw (2.0+0.08,-0.4) -- (2.0+0.08,-2);
\draw (2.0+0.16,-0.4) -- (2.0+0.16,-2);

\draw (1.6,0.16) -- (0.4,0.16);
\draw (1.6,0.08) -- (0.4,0.08);
\draw (1.6,0.0) -- (0.4,0.0);
\draw (1.6,-0.08) -- (0.4,-0.08);
\draw (1.6,-0.16) -- (0.4,-0.16);

\draw (2.0-0.16,0.4) -- (2.0-0.16,2);
\draw (2.0-0.08,0.4) -- (2.0-0.08,2);
\draw (2.0-0.0,0.4) -- (2.0-0.0,2);
\draw (2.0+0.08,0.4) -- (2.0+0.08,2);
\draw (2.0+0.16,0.4) -- (2.0+0.16,2);

\draw[fill=gray] (0,0) circle (0.38);

\draw (-0.28,-0.28) -- (-1.41,-1.41);
\draw (-0.28-0.055,-0.28+0.055) -- (-1.41-0.056,-1.41+0.055);
\draw (-0.28-0.110,-0.28+0.110) -- (-1.41-0.112,-1.41+0.110);
\draw (-0.28+0.055,-0.28-0.055) -- (-1.41+0.056,-1.41-0.056);
\draw (-0.28+0.110,-0.28-0.110) -- (-1.41+0.112,-1.41-0.110);

\draw (-0.28,0.28) -- (-1.41,1.41);
\draw (-0.28-0.055,0.28-0.055) -- (-1.41-0.055,1.41-0.055);
\draw (-0.28-0.11,0.28-0.11) -- (-1.41-0.11,1.41-0.11);
\draw (-0.28+0.055,0.28+0.055) -- (-1.41+0.055,1.41+0.055);
\draw (-0.28+0.11,0.28+0.11) -- (-1.41+0.11,1.41+0.11);

\node at (-1.65,-1.65) {$N$};
\node at (-1.65,1.65) {$N$};
\node at (2.0,-2.3) {$M$};
\node at (2.0,2.3) {$M$};

\draw[fill=gray] (2.0,0) circle (0.38);

\draw (2.4,0.16) -- (4,0.16);
\draw (2.4,0.08) -- (4,0.08);
\draw (2.4,0.0) -- (4,0.0);
\draw (2.4,-0.08) -- (4,-0.08);
\draw (2.4,-0.16) -- (4,-0.16);

\node at (4.35,0) {$2N$};

\endtikzpicture
\caption{
Global deformation (in the classification of \cite{Aharony:1997ju,Aharony:1997bh}) of the brane intersection shown in fig.~\ref{fig:5-pole-a},
corresponding to a relevant deformation of the dual SCFT.
\label{fig:5-pole-c}
}
\end{figure}

The asymptotic behavior for $M/N\rightarrow 0$ and $M/N\rightarrow\infty$ can be understood in more detail as follows.
For $M\rightarrow 0$, we expect the solution to reduce to a 3-pole configuration, since two of the residues in (\ref{eqn:5pole-residues}) vanish. Indeed, in that limit two of the zeros $s_n$ approach the real line and annihilate the poles $p_1$, $p_3$.
With one zero remaining in the interior of the upper half plane and three poles on the real line,
we indeed find a regular 3-pole configuration.
Correspondingly, the on-shell action as shown in fig.~\ref{fig:5-pole-b} for $M/N=0$ agrees 
with (\ref{eq:3-pole-I0}) evaluated with the remaining residues.
For large $M/N$, the behavior is not quite as immediately clear from the form of the residues.
But we can gain some intuition from looking at deformations of the web.
The solutions we are considering here describe the conformal phase of the dual SCFTs, where in the brane construction all external branes intersect at one point. 
Deformations of the web where the external branes are moved correspond to relevant deformations of the dual SCFT \cite{Aharony:1997ju,Aharony:1997bh}, and a particular example is shown in fig.~\ref{fig:5-pole-c}.
We may view it as gluing an intersection of $M$ NS5-branes and $2N$ D5-branes with an SL(2,$\mathds{R}$) rotated version of the ``N-junction''.
For large $M$, it suggests that the structure of the web is dominated by the intersection of $M$ NS5-branes and $2N$ D5-branes.
The number of degrees of freedom provided by the ``extra vertex'' compared to the 4-brane intersection of NS5 and D5-branes does not appear to scale with $M$, and we therefore expect the free energy of the 5-pole solution at large $M/N$ to approach the free energy of a 4-pole solution with charges corresponding to $M$ NS5  and $2N$ D5-branes.
As shown in fig.~\ref{fig:5-pole-b}, this is indeed the case.

\section{Entanglement entropy}\label{sec:EE}
In this section we use the Ryu-Takayanagi prescription \cite{Ryu:2006bv} to compute holographic entanglement entropies for the 5d SCFTs dual to the supergravity solutions. The main parts of the derivation will hold for a generic choice of the region for which we compute the entanglement entropy, as we will explain shortly, but our main interest is in regions of spherical shape.

The entanglement entropy is given by the area of a codimension-$2$ surface, anchored at a fixed time on the boundary of AdS$_6$ such that it coincides with the entangling surface. 
For a generic choice of entangling surface, we thus have to compute the area of an eight-dimensional surface $\gamma_8$ wrapping S$^2$ and $\Sigma$, and which is of codimension $2$ in AdS$_6$.
The resulting expression for the entanglement entropy reads
\begin{align}
 S_\mathrm{EE}&=\frac{\mathrm{Area}(\gamma_8)}{4G_\mathrm{N}}
 =\frac{1}{4G_\mathrm{N}}\int_{\gamma_8}\vol_{\gamma_8}~.
\end{align}
The volume form reduces to
\begin{align}
 \vol_{\gamma_8}&=f_6^4 f_2^2 \vol_{\gamma_4}\wedge \vol_{\mathrm{S}^2}\wedge\vol_\Sigma~,
\end{align}
where $\gamma_4$ is the codimension-2 minimal surface in a unit radius AdS$_6$ which is anchored at the conformal boundary and ends there on the entangling surface.
The computation of $S_\mathrm{EE}$ as a result simplifies to
\begin{align}\label{eq:EE-split}
 S_\mathrm{EE}&=\frac{1}{4 G_\mathrm{N}}\Vol_{\mathrm{S}^2}\cdot\mathcal I \cdot \mathrm{Area}(\gamma_4)~,
\end{align}
where $\mathrm{Area}(\gamma_4)$ is the area of the four-dimensional minimal surface in AdS$_6$
and with $g^{}_\Sigma=4\rho^2 |dw|^2$ we have
\begin{align}\label{eqn:cI-def}
\mathcal I &= 4 \int_\Sigma  d^2 w   f_6^4 f_2^2 \rho^2~.
\end{align}
The factor $4$ is a result of the ansatz (\ref{eqn:ansatz}) and we have $d^2w=dxdy$.
With the expressions for the metric functions in (\ref{eqn:metric}), we can further evaluate the integrand to find
\begin{align}\label{eq:I integ}
 \mathcal I&=\frac{8}{3} \int_\Sigma d^2 w \; \kappa^2\cG~.
\end{align}
We note in particular that, due to the factorization in (\ref{eq:EE-split}), once $\mathcal I$ is known the computation of entanglement entropies reduces to the analogous computation in AdS$_6$.

\subsection{Integrability near the poles}
We now show that even though the supergravity solution is singular at the poles $x=p_\ell$ on the boundary of $\Sigma$, the entanglement entropy is finite and does not receive contributions from the poles. To this end we use equation (\ref{eq:I integ}) together with the explicit expressions for $\kappa$ and $\cal G$ close to a pole derived in \cite{DHoker:2017mds}.
Namely, for $w=p_m+r e^{i \theta}$ we have 
\begin{subequations}\label{eq:near-pole}
\begin{align}
 \mathcal G &= 2 \kappa_m^2 r |\ln r| \sin \theta+\mathcal{O}(r^2 \ln r)~,
 &
 \partial_w\mathcal G&= i\kappa_m^2\ln r +\mathcal O(r\ln r)~,
\end{align}
and
\begin{align}
 \kappa^2 &= \kappa_m^2 \frac{\sin\theta}{r}+\mathcal{O}(r^0)~,
\end{align}
\end{subequations}
where 
\begin{align}
\kappa_m^2&=2 i \sum_{\ell\neq m}^{}\frac{Z^{[\ell m]}}{p_m-p_\ell}~.
\end{align}
This implies that the integrand of $\cal I$ close to the pole behaves as  $\mathcal O(r|\ln r|)$, which is integrable.
Moreover, we see that, like in the direct computation of the free energy in sec.~\ref{sec:on-shell-action}, we can introduce a cut-off around the poles and evaluate the integrals, and removing the cut-off does not yield localized contributions from the poles.

\subsection{Explicit evaluation}

We now turn to a more explicit evaluation of the integral $\mathcal I$ given in (\ref{eq:I integ}). 
We can use the fact that
\begin{align}
 \kappa^2&=-\partial_w \partial_{\bar w}\cG~,
\end{align}
to integrate by parts. Namely, using $\kappa^2\cG= -\partial_w (\cG\partial_{\bar w}\cG)+(\partial_{\bar w}\cG)\partial_w\cG$.
From the near-pole expansions in eq.~(\ref{eq:near-pole}), we see that $\cG\partial_{\bar w}\cG$ goes to zero not only at generic points of the boundary, but also at the poles.
The boundary contribution therefore vanishes and we find
\begin{align}\label{eq:cI-2}
 \mathcal I&=\frac{8}{3}\int_\Sigma d^2w (\partial_{\bar w}\cG)\partial_w\cG~.
\end{align}
The generic form of $\partial_w\cG$ can be obtained straightforwardly from (\ref{eq:kappa-G}) and yields
\begin{align}
 \partial_w \cG&=(\bar\cA_+-\cA_-)\partial_w\cA_+  + (\cA_+-\bar\cA_-)\partial_w\cA_-~.
\end{align}
Evaluating this explicitly using the regularity conditions (\ref{eqn:constr}) yields
\begin{align}
 \partial_w \cG&= \sum_{\substack{\ell,k=1 \\ \ell\neq k}}^L Z^{[\ell k]}  \ln \left|\frac{w-p_\ell}{p_k-p_\ell}\right|^2\frac{1}{w-p_k}~.
\end{align}
This relation allows us to write $\cal I$ explicitly as
\begin{align}\label{eq:EE}
\mathcal I &=-\frac{8}{3}  \sum_{\substack{\ell,k,m,n=1 \\ \ell\neq k, m\neq n}}^L Z^{[\ell k]}  Z^{[mn]}\int_\Sigma d^2 w   \ln\left|\frac{w-p_\ell}{p_k-p_\ell}\right|^2 \ln\left|\frac{w-p_m}{p_m-p_n}\right|^2 \frac{1}{\bar w-p_n}\frac{1}{w-p_k}~.
\end{align}
This expression becomes manifestly real upon symmetrizing the integrand under the exchange of the index pairs $(\ell,k)$ and $(m,n)$, which are independently summed over. In addition, using charge conservation, one can show that the combination $dw \partial_w\mathcal{G}$   is invariant under SL(2,$\mathbb{R}$) transformations
\begin{align}
w&\to \frac{aw+ b}{cw+d}~, & p_k&\to \frac{ap_k+b}{cp_k +d}~,
\end{align}
with $ad-bc=1$. The expression for $\mathcal I$ in (\ref{eq:cI-2}) is therefore SL(2,$\mathbb{R}$) invariant, as expected, and we can again fix the location of three poles at arbitrary positions.

\subsection{Spherical regions}
For the specific case of a spherical entangling surface   of radius $r_0$ at a fixed $t=t_0$, 
we just have to evaluate the area of the corresponding minimal surface in an AdS$_6$ of unit radius.
We choose coordinates in AdS$_6$ such that
\begin{align}
 ds^2_{\mathrm{AdS}_6}&= \frac{dz^2 -dt^2+  dr^2  + r^2 d\Omega_{S^3}^2}{z^2}~.
\end{align}
The minimal surface can be parametrized by $r=r(z)$ and its area is given by
\begin{align}
 \mathrm{Area}(\gamma_4)&=\Vol_{\mathrm{S}^3}\int dz  \frac{r(z)^3\sqrt{1+  r'(z)^2}}{z^4}~.
\end{align}
Extremizing this functional yields the usual solution
\begin{align}
r(z)&=\sqrt{r_0^2-z^2}~.
\end{align}
The $z$ integral is divergent at $z=0$, and the choice of cut-off follows the same logic as outlined for the free energy in appendix \ref{sec:ads6vol}. With a bulk IR/field theory UV cutoff at $z=\epsilon$, the integral becomes
\begin{align}
\int^{r_0}_\epsilon dz  \frac{r(z)^3}{z^4} \sqrt{1+  r'(z)^2} &= \frac{r_0^3}{3\epsilon^3}- \frac{r_0}{\epsilon} + \frac{2}{3} + \mathcal O(\epsilon)~.
\end{align}
Although holographic renormalization for submanifolds is well understood \cite{Graham:1999pm}, the divergences in the entanglement entropy are usually kept, as a reflection of the short-distance behavior of QFTs.
The universal part in odd dimensions, however, is the finite contribution and for the surfaces considered here given by
\begin{align}\label{eq:area-ren}
 \mathrm{Area}_\mathrm{ren}(\gamma_4)&=\frac{2}{3}\Vol_{\mathrm{S}^3}~.
\end{align}

In summary, the entanglement entropy for a spherical region is given by the expression in (\ref{eq:EE-split}), with the universal part of the area of the minimal surface in (\ref{eq:area-ren}) and $\mathcal I$ given in (\ref{eq:EE}).
We note that this expression manifestly exhibits the same scaling with the residues $Z_+^\ell$, corresponding to the charges of the external 5-branes, as the expression for the on-shell action in (\ref{eq:on-shell-action}).

\subsection{Matching to free energy}\label{sec:EE-merge}
In this section we show that for all the examples discussed in sec.~\ref{subsec:3-4-5 poles action} the finite part of the holographic entanglement entropy for a spherical region is equal to minus the finite part of the free energy on S$^5$. To accomplish this we will reduce part of the two-dimensional integral over $\Sigma$ appearing in equation (\ref{eq:I integ}) to a one-dimensional integral over the real line which has the same form as the one-dimensional integral appearing in the on-shell action (\ref{eq:on-shell-action}), and show that the remaining part vanishes.

Using $\kappa^2=-\partial_w\partial_{\bar w}\cG$ and the definition of $\cG$ in (\ref{eqn:Gdef}),
the integral $\mathcal I$ given in (\ref{eq:I integ}) can be rewritten as 
\begin{align}
 \mathcal I &=-\frac{8}{3} \int_\Sigma d^2w \:  \partial_w \partial_{\bar w} \mathcal G \left(|\cA_+|^2-|\cA_-|^2+\mathcal{B}+\bar{ \mathcal{B}}\right)~.
\end{align}
We split $\mathcal{I}$ into two terms:
\begin{subequations}
\begin{align}
\mathcal{I}&=\mathcal{I}_1+\mathcal{I}_2~,\\
\mathcal{I}_1&=-\frac{4}{3} \int_\Sigma d^2w \:  \partial_w \partial_{\bar w} \mathcal G \left(\mathcal{B}+\bar{ \mathcal{B}}\right)~, \\
\mathcal{I}_2&=-\frac{8}{3} \int_\Sigma d^2w \:   \partial_w \partial_{\bar w} \mathcal G \left(|\cA_+|^2-|\cA_-|^2+\frac{1}{2}\left(\mathcal{B}+\bar{ \mathcal{B}}\right)\right)~.
\label{eq:I2exp}
\end{align}
\end{subequations}
First we evaluate  $\mathcal{I}_1$ and will argue below that the second integral $\mathcal{I}_2$ vanishes. 
Since $\mathcal B$ is holomorphic, we can write the integrand of $\mathcal I_1$ as a sum of total derivatives
\begin{align}
 \partial_w \partial_{\bar w} \mathcal G \left(\mathcal{B}+\bar{ \mathcal{B}}\right)
 &=\partial_{\bar w}\left(\partial_w\cG (\cB+\bar \cB)\right)-\partial_w\left(\cG \partial_{\bar w}\bar{\cB}\right)~.
\end{align}
The boundary term resulting from the second term vanishes since $\cG=0$ on $\partial\Sigma$. 
Switching to real coordinates we therefore find
\begin{equation}
\mathcal{I}_1=\frac{2i}{3} \int_{-\infty}^{\infty} dx\: \partial_w \mathcal{G}\left(\mathcal{B}+\bar{\mathcal{B}}\right)\bigg|_{y=0}~.
\end{equation}
To evaluate the integrand we use that $\cG=0$ on the real line and hence $\cB+\bar\cB=-|\cA_+|^2+|\cA_-|^2$. 
This yields
\begin{align}
\partial_w \mathcal{G}\Big|_{y=0}&=2 \sum_{\substack{m,n=1\\ m \neq n}}^L \frac{Z^{[mn]}}{x-p_n} \ln\bigg|\frac{x-p_m}{p_m-p_n}\bigg|~,\\
\mathcal{B}+\bar{\mathcal{B}}\Big|_{y=0}&=2 \pi i\sum_{\substack{\ell,k=1\\ k\neq \ell}}^L Z^{[\ell k]} \ln\bigg|\frac{x-p_k}{p_k-p_\ell}\bigg|\Theta\left(p_\ell-x\right)~.
\end{align}
Thus we get
\begin{align}
\mathcal{I}_1&=-\frac{8 \pi }{3}\sum_{\substack{\ell,k,m,n=1 \\ \ell\neq k, m\neq n}}^L\int_{-\infty}^{\infty} dx \frac{Z^{[\ell k]}{ Z^{[m n]}}}{x-p_n} \ln\bigg|\frac{x-p_m}{p_m-p_n}\bigg|\ln\bigg|\frac{x-p_k}{p_k-p_\ell}\bigg|\Theta\left(p_\ell-x\right)~.
\end{align}
Plugging this result into (\ref{eq:EE-split}) gives the following contribution to the entanglement entropy
\begin{align}
S_{\mathrm{EE}1}\label{eq:see1}
&= -\frac{4\pi }{9 G_{N}}\mathrm{Vol}_{\mathrm{S}^2}\mathrm{Vol}_{\mathrm{S}^3} \sum_{\substack{\ell,k,m,n=1 \\ \ell\neq k, m\neq n}}^L Z^{[\ell k]}{ Z^{[ m n]}} \int_{-\infty}^{p_\ell} dx  \ln\bigg|\frac{x-p_m}{p_m-p_n}\bigg|\ln\bigg|\frac{x-p_k}{p_k-p_\ell}\bigg|\frac{1}{x-p_n}~.
\end{align}
We can compare this result with the value of the finite part of the on-shell action derived in section \ref{subsec:on-shell-action integral}:
\begin{equation}
( S_\mathrm{IIB}^\mathrm{E})^{\text{finite}}=\frac{8 }{{9} G_\mathrm{N}} \mathrm{Vol}_{\mathrm{S}^5} \mathrm{Vol}_{\mathrm{S}^2}
 \sum_{\substack{\ell,k,m,n=1 \\ \ell\neq k, m\neq n}}^L Z^{[\ell k]}Z^{[m n]}
 \int_{-\infty}^{p_\ell}dx\,\ln\left|\frac{x-p_k}{p_\ell-p_k}\right| \,\ln\left|\frac{x-p_m}{p_m-p_n}\right| \frac{1}{x-p_n}~.
\end{equation}
Inserting the expressions for the volumes of the 2-, 3- and 5-sphere given by
\begin{equation}
\Vol_{\mathrm{S}^2}=4\pi~,  \quad\quad
\Vol_{\mathrm{S}^3}= 2\pi^2~, \quad\quad
\Vol_{\mathrm{S}^5} = \pi^3~,  
\end{equation}
confirms the equality  of  the finite parts of the entanglement entropy and the on-shell action  \begin{equation}
(S_{\mathrm{EE}1})^{\mathrm{finite}}=-( S_\mathrm{IIB}^\mathrm{E})^{\text{finite}}~.
\end{equation}

What remains to be shown is that  the integral $\mathcal{I}_2$ vanishes and hence $S_{\mathrm{EE}1}$ given in (\ref{eq:see1}) is the  complete expression for the  finite part of the entanglement entropy. The integral  $\mathcal{I}_2$ given in (\ref{eq:I2exp}) 
can be rearranged as follows
\begin{align}
\mathcal I_2&=-\frac{4}{3}\int_{\rm \Sigma} d^2w \;(\mathcal{G}+|\cA_+|^2-|\cA_-|^2)\partial_{w}\partial_{\bar w}\mathcal{G}
\\
&=-\frac{4}{3}\int_{\rm \Sigma} d^2w 
\left( -\partial_w \mathcal{G} \partial_{\bar w} \mathcal{G}+\partial_{w}\partial_{\bar w} \mathcal{G}(|\cA_+|^2-|\cA_-|^2) \right)~.
\end{align}
Using the explicit expressions of $\cA_+$ and $\cA_-$ we get:
\begin{align}
 \mathcal I_2=-\frac{4}{3}\sum_{\substack{\ell,k,m,n=1 \\ \ell\neq k, m\neq n}}^LZ^{[ m n]} Z^{[\ell k]} 
\int_{\rm \Sigma} d^2w\,\frac{1}{\bar w-p_m}\bigg(
&\ln\left|\frac{w-p_\ell}{p_k-p_\ell}\right|^2\ln\left|\frac{w-p_n}{p_m-p_n}\right|^2\frac{1}{w-p_k}
\nonumber\\&\hskip 0.5in
+\ln\frac{w-p_\ell}{|p_k-p_\ell|}\ln\frac{\bar w-p_k}{|p_k-p_\ell|}\frac{1}{w-p_n}\bigg)~.
\end{align}
For the three-pole solutions we have shown analytically that this term vanishes, and for the four and five pole solutions discussed in sec.~\ref{subsec:3-4-5 poles action} we have verified this numerically. For all these cases we therefore find that the finite parts of the entanglement entropy and the on-shell action are related as expected on general grounds \cite{Casini:2011kv}.
Although we do not currently have an analytic proof, this certainly suggests that the relation between free energy and entanglement entropy holds for all the solutions reviewed in sec.~\ref{sec:review}.

\section{Discussion}\label{sec:discussion}

We have studied the free energy  of the field theories described by the supergravity solutions constructed in \cite{DHoker:2016ysh,DHoker:2017mds}. 
Unlike for previously known AdS$_6$ solutions in type IIA supergravity, the computation of the free energy is straightforward albeit technically non-trivial for these solutions. We conclude that the isolated singularities that are present are mild and do not obstruct holographic computations.
Moreover, the computation of the free energy via the entanglement entropy of a spherical region reproduces the result of the direct computation, a relation which is expected to hold on general grounds but corresponds to non-trivial integral identities in the explicit solutions considered here.
These results support the interpretation of the solutions as holographic duals to the five-dimensional superconformal field theories engineered in type IIB string theory via 5-brane webs, and give first quantitative indications on the nature of the dual field theories.
We will close with a more detailed discussion of the implications and some directions for future research.

An immediate question concerning the supergravity solutions and their interpretation concerns the external 5-branes. In  \cite{DHoker:2016ysh,DHoker:2017mds} the singularities located at the poles were interpreted as the remnants of the external $(p,q)$ five branes in the brane web construction of the five dimensional field theories,  which flow to the dual SCFT in the conformal limit.  

Whether brane webs with parallel external branes lead to well-defined five-dimensional SCFTs was initially questioned, with one potential obstacle being light states on the parallel branes that may not decouple from the field theory on the intersection. It was later argued that these light states do in fact decouple \cite{Bergman:2012kr}, and webs with parallel external branes indeed lead to well-defined 5d field theories after factoring out the decoupled states \cite{Bergman:2013ala,Bao:2013pwa,Hayashi:2013qwa,Taki:2013vka,Taki:2014pba}.
       
For our supergravity solutions this immediately poses the question of whether or not they include contributions from parallel external branes, e.g.\ in the form of states localized around the poles on $\partial\Sigma$.
The computation of the free energy in sec.~\ref{sec:on-shell-action} and \ref{sec:EE} indicates that this may not be the case:
In both cases we could introduce a cut-off around the poles on $\Sigma$ and effectively remove them from the geometry.
If states localized around the poles would contribute, we would expect the free energy to change by a finite amount,
i.e.\ we would expect to produce non-trivial boundary terms.
The scaling analysis for  both cases shows that this is not the case, and we therefore do not seem to see contributions from the external 5-branes.

Another open question about the solutions was whether and how the external 5-branes end on 7-branes.
We discussed in \cite{DHoker:2017mds} that there was no indication for the presence of 7-branes and that a natural expectation would be that the supergravity solutions describe brane webs with only 5-branes.
But with only access to the intersection, the possibility that external 7-branes would just not be directly accessible from the supergravity solution remained a valid option.
Another natural option could then be that all 5-branes within a given stack of external 5-branes end on the same 7-brane.
This allows for a brane web realization of the USp($N$) theory which was initially engineered in type IIA string theory \cite{Bergman:2012kr}.
Our results for the free energy and entanglement entropy, however, disfavor this option:
The scaling of the free energy in the USp($N$) theory is $N^{5/2}$, which is different from the scaling in the 4-pole type IIB supergravity solutions discussed in sec.~\ref{sec:on-shell-action}.
In particular, the solutions with $Z_+^1=-Z_+^3=(1+i)N$ and $Z_+^2=-Z_+^4=(-1+i)N$, if all external branes within a given stack would end on the same 7-brane, would realize the USp($N$) theory.
But the scaling we find is $N^4$ instead of $N^{5/2}$.
This suggests that the dual SCFTs may rather be of the long quiver type, as discussed in \cite{DHoker:2017mds}.
       
Finally, our results for the free energy in specific examples provide a clear target for field-theory computations. 
The overall factor of $\zeta(3)$ in all the examples discussed in sec.~\ref{subsec:3-4-5 poles action} 
provides a hint on the solution for the eigenvalue distribution of the matrix model resulting from supersymmetric localization in the field theory. E.g., the matrix model action derived in \cite{Kallen:2012va} involves explicit factors of $\zeta(3)$, along with polylogarithms, which gives an indication on where the eigenvalue distribution has to have support for these terms to play a role. We also note that the terms of interest dropped out of the calculations in \cite{Jafferis:2012iv}, precisely due to the properties of the eigenvalue distribution for these theories.

\begin{acknowledgments}
We are happy to thank Oren Bergman and Diego Rodriguez-Gomez for many insightful discussions.
We also acknowledge the Aspen Center for Physics, which is supported by National Science Foundation grant PHY-1066293, for hospitality during the workshop ``Superconformal Field Theories in $d\geq 4$'' and thank the organizers and participants for the enjoyable and inspiring conference.
The work of all four authors  is supported in part by the National Science Foundation under grant   PHY-16-19926.  
\end{acknowledgments}

\appendix

\section{Type IIB on-shell action as boundary term}\label{app:action-boundary-term}
To recall how the type IIB supergravity action can be written as a boundary term on-shell, we start from the action in the form \cite{Polchinski:1998rr}
\begin{align}\label{eq:A1}
S_{\mathrm{IIB}}&=\frac{1}{2 \kappa^2} \int d^{10}x \sqrt{-g}\left(R-\frac{\partial_\mu \bar \tau \partial^\mu  \tau}{2 \left(\text{Im}\tau\right)^2}-\frac{\mathcal M_{ij}}{2} F_3^i \cdot F_3^j-\frac{1}{4}|\tilde F_5|^2\right)-\frac{\epsilon_{ij}}{8 \kappa^2}\int C_4 \wedge F_3^i\wedge F_3^j~,
\end{align}
where the dot product is defined as $Q_{p}\cdot F_{p}=\frac{1}{p!} g^{\mu_1 \nu_1}...g^{\mu_p \nu_p}Q_{\mu_1...\mu_p}F_{\nu_1...\nu_p}$
and $\kappa^2=8\pi G_\mathrm{N}$ with Newton's constant $G_\mathrm{N}$.
In the main part we will not use the short hand $\kappa^2$ to avoid confusion with the composite quantity defined in (\ref{eq:kappa-G}).
The field strengths are defined as 
\begin{align}
F_3^i&=dC_2^i~,&
F_5&=dC_4~,
&
\tilde F_5&=F_5-\frac{1}{2}C_2^2\wedge F_3^1+\frac{1}{2}C_2^1\wedge 
F_3^2,
\end{align}
with $i=1,2$ where $i=1$ and $i=2$ correspond to the NS-NS and R-R 2-forms, respectively.
The $2\times 2$ matrix $\mathcal M$ is given by
\begin{align}
 \mathcal M_{ij}&=\frac{1}{\text{Im} \tau} \begin{bmatrix}
|\tau|^2 & -\text{Re}\tau  \\
-\text{Re}\tau & 1
\end{bmatrix}~.
\end{align}
As shown in \cite{Okuda:2008px}, on-shell this action reduces to a boundary term. 
For $C_4=0$, which applies for all configurations considered here, this boundary term reduces to 
\begin{align}\label{eqn:boundary-action1}
S_{\mathrm{IIB}}&= \frac{1}{2 \kappa^2}\int d\left(-\frac{1}{4}\mathcal M_{ij}C_2^i\wedge \star F_3^j\right)~.
\end{align}

To translate this expression to our conventions for the supergravity fields, we 
combine the two real 2-forms $C_2^i$ into one complex 2-form, $C_2=C_2^1+i C_2^2$, 
with field strength $F_3=dC_2$, and redefine the fields as follows,
\begin{align}
B&=\frac{1+i \tau}{1-i \tau}~,
&
f^2&=\left(1-|B|^2\right)^{-1}~.
\end{align} 
In terms of $f$, $B$ and $C_2$, and eliminating $\kappa^2$ in favor of $G_\mathrm{N}$, the boundary term (\ref{eqn:boundary-action1}) becomes
\begin{align}\label{eq:A3}
 S_{\mathrm{IIB}}=-\frac{1}{64\pi G_\mathrm{N}}\int d \bigg(\frac{1}{2}f^2 (1+|B|^2)&\left(\bar C_2\wedge \star dC_2+C_2\wedge \star d\bar C_2\right)
 \nonumber\\&
 -f^2 \bar B C_2\wedge \star dC_2-f^2 B \bar C_2 \wedge \star d\bar C_2\bigg)~.
\end{align}
For the configurations we are interested in, there is no non-trivial dependence on the AdS$_6$ coordinates.
We can therefore Wick rotate between Lorentzian and Euclidean signature purely within the AdS$_6$ part, which 
only enters through the volume form and at most accounts for a sign in the on-shell action.
That sign can be fixed directly in Euclidean signature, where we want
$\mathcal Z=\int Dg\exp(-S)$ with $S$ positive semi-definite, such that $F=-\ln \mathcal Z$ is non-negative.
For AdS there are the usual subtleties with divergences and holographic renormalization, and we will discuss this in more detail in app.~\ref{sec:ads6vol}.
We will demand the leading divergent term in the regularized free energy to be positive, and this corresponds to
\begin{align}\label{finresappa}
S_\mathrm{IIB}^\mathrm{E}=\frac{1}{64\pi G_\mathrm{N}}\int d \left(\frac{1}{2}f^2 (1+|B|^2)\,\bar C_2\wedge \star dC_2-f^2 \bar B C_2\wedge \star dC_2+\mathrm{c.c.}\right)~.
\end{align}

\section{Holographic renormalization}\label{sec:ads6vol}
Holographic renormalization of the gravity theory on an asymptotically-AdS space becomes considerably more involved if the geometry does not reduce to a simple product form in the near-boundary limit.
In general, the entire ten-dimensional geometry has to be considered with a nine-dimensional cut-off surface limiting the range of the radial coordinate in the asymptotic part of the geometry.
There is a substantial amount of freedom in choosing this cut-off surface, which by the usual AdS/CFT lore corresponds to the freedom to choose a regularization scheme on the field theory side.
In many cases one can restrict the choice of the cut-off surface by symmetry requirements.
E.g., for AdS$_5\times$S$^5$, one would require the cut-off surface to respect the S$^5$ isometries, which essentially reduces the problem of finding counterterms to the AdS$_5$ factor.
For our geometries the analogous symmetry argument restricts the location of the cut-off on the AdS$_6$ radial coordinate to be independent of the location on S$^2$. The dependence on the location on $\Sigma$, however, is not restricted by that requirement.

For definiteness, we will choose global coordinates on Euclidean AdS$_6$ such that the metric takes the form
\begin{equation}\label{eq:A6}
g_{\mathrm{AdS}_6}=du^2+\sinh u^2 g_{\mathrm{S}^5}~,
\end{equation}
with $u\in[0,\infty)$.
The cut-off surface should provide an upper bound on the range of $u$.
The perhaps most natural choice is to pick a small $\epsilon\in\mathds{R}^+$ and require $u<\arcsinh(1/\epsilon)$.
The cut-off surface is then the nine-dimensional surface defined by $u=\arcsinh(1/\epsilon)$.
This regulator is invariant under the isometries of the S$^5$ inside AdS$_6$ and under the isometries of S$^2$, corresponding to spacetime isometries and R-symmetry in the dual field theory, respectively.
However, any cut-off surface of the form $u=\arcsinh(1/\epsilon(w))$, with $\epsilon(w)$ small throughout $\Sigma$, satisfies these requirements as well, and we are indeed free to choose any of them.

The value of the regularized on-shell action will certainly depend on the choice of regulator, as it usually does. The freedom in choosing a cut-off surface is enhanced here compared to the simpler cases with highly symmetric bulk geometries (where the freedom essentially boils down to rescalings of the cut-off), but the fact that there is ambiguity is by no means a new feature of the solutions considered here.
More importantly, after proper holographic renormalization the universal parts of any physical quantity considered still have to be independent of the choice of regulator.
We can therefore pick the simplest one, where $\epsilon$ is constant over $\Sigma$, as long as we only ask for physically meaningful (universal) quantities.

Moreover, since we have an even-dimensional AdS space with odd-dimensional field theory, there are no finite counterterms from the metric sector: the volume form on AdS$_6$ scales like $\epsilon^{-5}$, and all other covariant quantities constructed from the induced metric  on the cut-off surface (including the GHY term) have an expansion in even powers of $\epsilon$.
That means the covariant boundary terms scale as odd powers of $\epsilon$ and do not produce finite contributions.
We have not explicitly verified that this holds for the other fields as well, but since they are related by supersymmetry we expect the corresponding covariant counterterms to scale with odd powers of $\epsilon$ as well.
There is therefore no ambiguity in choosing a renormalization scheme, and we can read off the universal part directly, e.g.\ as the finite part of the free energy, without going through the proper procedure of holographic renormalization.

With the cut-off $u<\arcsinh(1/\epsilon)$, the holographic renormalization indeed reduces to a pure AdS$_6$ problem, with the regularized volume of AdS$_6$ given by 
\begin{align}
 \Vol_{\mathrm{AdS}_6}&= \Vol_{\mathrm{S}^5}\int_0^{\arcsinh\frac{1}{\epsilon}}  \sinh^5 u du
 =\Vol_{\mathrm{S}^5}\left( \frac{1}{5 \epsilon ^5}-\frac{1}{6 \epsilon ^3}+\frac{3}{8 \epsilon }-\frac{8}{15}+\mathcal O \left({\epsilon}\right)\right)~.
\end{align}
As argued above, the universal part can be extracted immediately and is given by
\begin{align}
 \Vol_{\mathrm{AdS}_6,\mathrm{ren}}&= 
 -\frac{8}{15}\Vol_{\mathrm{S}^5}~.
\end{align}

\bibliography{5dee}
\end{document}